\documentclass[12pt]{article}
\pdfoutput=1
\usepackage[margin=3cm]{geometry}
\usepackage[utf8]{inputenc}
\usepackage{amsmath}
\usepackage{amssymb}
\usepackage{authblk} 
\usepackage{xcolor}
\usepackage{graphicx}
\usepackage{float}
\usepackage{tikz}
\usetikzlibrary{calc}
\usepackage{hyperref}    
\usepackage{cleveref}     

\Crefname{table}{Table}{Tables}
\crefname{figure}{figure}{figures}
\Crefname{figure}{Figure}{Figures}

\definecolor{tealblue}{rgb}{0.21, 0.56, 0.63}
\hypersetup{colorlinks=true,allcolors = tealblue,linktocpage=true}

\usepackage{pgf}

\usepackage[backend=bibtex, sorting=none,s tyle=phys, eprint=true, doi=false, biblabel=brackets]{biblatex}
\bibliography{bibliography}
\numberwithin{equation}{section}

\newenvironment{eqaed}
    {\begin{equation}
    \begin{aligned}
    }
    { 
    \end{aligned}
    \end{equation}
    \ignorespacesafterend
    }

\begin{document}
\vspace*{-1.5cm}
\begin{flushright}
  {\small
  MPP-2026-154\\
  }
\end{flushright}
\title{A LooKK at the Higuchi Bound}

\date{}

\author{Dieter Lüst$^{a,b}$, Joaquin Masias$^{a}$}
\affil{${}^a$\emph{Max-Planck-Institut f\"ur Physik (Werner-Heisenberg-Institut)}\\ \emph{Boltzmannstraße 8, 85748 Garching, Germany}}

\affil{${}^b$\emph{Arnold Sommerfeld Center for Theoretical Physics, Ludwig-Maximilians-Universität München, 80333 München, Germany}}

{\let\newpage\relax\maketitle}
\begin{abstract}
Massive spin-2 fields on de Sitter must satisfy the Higuchi bound, $m^2\geq 2H^2$. Compactifications of higher-dimensional gravity to $\mathrm{dS}_4$ come with a tower of Kaluza-Klein gravitons whose masses are fixed by the internal geometry, so the bound becomes a constraint on the compactification. We propose that the KK gravitons of a consistent compactification never violate it, and test this in a few examples. On a circle stabilized by Casimir energy the bound holds unless the circle shrinks below the species scale, and on a warped interval with negative tension branes the tower is gapped at $m^2\geq \tfrac{9}{4}H^2$ for any length of the interval. On group manifolds, we argue that if the internal curvature is not parametrically larger than the Hubble scale, the bound can apparently be violated by smooth deformations. The violation disappears once one promotes these deformations to moduli, which are immediately extremized when solving the 10d equations of motion.
\end{abstract}
\newpage
\section{Introduction} \label{sec:introduction}

De Sitter space is the maximally symmetric, positively curved manifold in Lorentzian signature, obtained in Einstein gravity from a positive cosmological constant $\Lambda = 3H^2$, with the Hubble parameter $H$ setting its characteristic scale. It is central to cosmology, as a quasi-de Sitter phase of accelerated expansion is the preferred description for the inflationary early universe \cite{Guth:1980zm, Linde:1981mu, Planck:2018jri, BICEP:2021xfz}, while the late universe is governed by dark energy, long considered to be a cosmological constant \cite{Weinberg:1988cp,Planck:2018vyg} (although recent observations suggest it might instead be dynamical \cite{DESI:2025zgx, Sabogal:2025jbo, Anchordoqui:2026hys}). On a de Sitter background, unitarity places a lower bound on the mass of bosonic fields in massive representations \cite{Higuchi:1986py}. The bound is trivial for spins 0 and 1, but for spin-2 it requires $m^2_{\text{spin-2}}\geq 2 H^2$, which is the Higuchi bound. Lighter masses propagate non-unitarily, so light massive gravitons cannot exist on a $\mathrm{dS}$ background, as unitary spin-2 irreps of $SO(1,4)$ do not extend below $m^2=2H^2$ \cite{Higuchi:1986wu}.

Whether string theory can realize consistent, under control $\mathrm{dS}$ extrema at all is a long standing question \cite{Kachru:2003aw, Conlon:2005jm} (see \cite{Baumann:2014nda} for reviews). Superstring theories are only consistent in ten dimensions, so six of them must be compactified, and dimensional reduction from $4+p$ to $4$ dimensions inevitably produces a tower of massive Kaluza-Klein gravitons, the lightest of which typically has mass $m_{\rm KK}\simeq \mathcal{V}^{-1/p}$, in the absence of warping or large sub-cycles, with $\mathcal{V}$ the volume of the compact $p$-manifold. The Higuchi bound must then hold for the entire tower, and in particular for the lightest mode. Since $m_{\rm KK}$ is fixed by the compactification data rather than being a free parameter, and since $H$ is itself fixed by the four dimensional vacuum energy, which depends on the internal manifold, the Higuchi bound becomes a constraint on the internal geometry, and it is not obvious a priori that it must hold.

Superstring theories are only consistent in ten dimensions, so six of them must be compactified. Dimensional reduction from $4+p$ to $4$ dimensions inevitably produces a tower of massive Kaluza-Klein gravitons, the lightest of which typically has mass $m_{\rm KK}\simeq \mathcal{V}^{-1/p}$, in the absence of warping or large sub-cycles, with $\mathcal{V}$ the volume of the compact $p$-manifold. The Higuchi bound must then hold for the entire tower, and in particular for the lightest mode. Since neither $m_{\rm KK}$ nor $H$ is a free parameter, as both depend on the compactification data, this becomes a constraint on the internal geometry, and it is not obvious a priori that it must hold.

A similar tension was pointed out in \cite{Lust:2019lmq}, where the tower of higher spin string excitations was argued to be in principle incompatible with the bound, so that these modes admit no valid effective description below the Hubble scale (see also \cite{Noumi:2019ohm,Scalisi:2019gfv,Kato:2021rdz}), and the bound has been used along the same lines to constrain light spin-2 fields \cite{Klaewer:2018yxi, Lust:2019zwm} and slow-roll inflation \cite{Kleban:2015daa,Luben:2020wim, Anchordoqui:2022svl, Anchordoqui:2026edd}. In this paper we consider instead Kaluza-Klein modes. We study such towers on geometries of the form $\mathrm{dS}_4 \times X$, with $X$ compact, and propose that the KK gravitons of a consistent compactification can never violate the bound. We then test the proposal in several examples. In all the cases considered the bound is either satisfied with $m_{\rm KK}^2>2 H^2$, or its violation signals that the underlying theory is not self-consistent.

\section{Massive spin-2 on \texorpdfstring{$\mathrm{dS}_4$}{dS4}} \label{sec:ii}

Consider a symmetric tensor field $h_{\mu\nu}$ on a fixed $\mathrm{dS}_4$ background with metric $\bar g_{\mu\nu}$ and Hubble parameter $H$. The background satisfies
\begin{equation}
\bar R_{\mu\nu} = 3H^2 \bar g_{\mu\nu}, 
\qquad 
\bar R_{\mu\nu\rho\sigma} = H^2\left(\bar g_{\mu\rho}\bar g_{\nu\sigma} - \bar g_{\mu\sigma}\bar g_{\nu\rho}\right).
\end{equation}
The Fierz-Pauli action \cite{Fierz:1939ix, deRham:2014zqa} on a curved background can be written as
\begin{equation}
S_{\text{FP}} = \frac{M_{\rm Pl}^2}{8}\int d^4x \sqrt{-\bar g}\left[\,
h^{\mu\nu}\,\mathcal{E}_{\mu\nu}{}^{\rho\sigma}\,h_{\rho\sigma}
-\,
m^2\left(h_{\mu\nu}h^{\mu\nu}-h^2\right)\right],
\label{eq:FPaction}
\end{equation}
where $h \equiv \bar g^{\mu\nu}h_{\mu\nu}$ and 
\begin{eqaed}
\mathcal{E}_{\mu\nu}{}^{\rho\sigma} h_{\rho\sigma}
&\equiv
-\bar\nabla^2 h_{\mu\nu}
-\bar\nabla_\mu \bar\nabla_\nu h
+\bar\nabla_\mu \bar\nabla^\rho h_{\rho\nu}
+\bar\nabla_\nu \bar\nabla^\rho h_{\rho\mu}
-\bar g_{\mu\nu}\!\left(\bar\nabla_\rho \bar\nabla_\sigma h^{\rho\sigma}-\bar\nabla^2 h\right)
\\
&\quad
-2\,\bar R_{\mu\rho\nu\sigma}\,h^{\rho\sigma}
+\bar R_\mu{}^{\rho} h_{\rho\nu}
+\bar R_\nu{}^{\rho} h_{\rho\mu}
-\bar R_{\mu\nu}\,h
-\bar g_{\mu\nu}\,\bar R_{\rho\sigma} h^{\rho\sigma}
+\bar R\,h_{\mu\nu}
-\frac{1}{2}\bar g_{\mu\nu}\bar R\,h .
\end{eqaed}
It is most convenient to write the kinetic term in terms of the Lichnerowicz operator \cite{Lichnerowicz:1961}, acting on a symmetric tensor $t_{\mu\nu}$ as
\begin{equation}
\Delta_L t_{\mu\nu} \equiv
-\bar\nabla^2 t_{\mu\nu}
-2\bar R_{\mu\rho\nu\sigma} t^{\rho\sigma}
+2\bar R_{(\mu}{}^{\rho} t_{\nu)\rho}.
\label{eq:LichnerowiczDef}
\end{equation}
For transverse-traceless tensors, $\bar\nabla^\mu t_{\mu\nu}=0$ and $t=0$, and for $\mathrm{dS}_4$ one finds
\begin{equation}
\Delta_L t_{\mu\nu} = \left(-\bar\nabla^2 + 8H^2\right)t_{\mu\nu}.
\label{eq:LichnerowiczTT}
\end{equation}
Here $\bar\nabla^2$ is the raw Laplacian with respect to the background metric.
The linearized equations of motion for the massive transverse-traceless modes then take the standard form
\begin{equation}
\left(\Delta_L - 6H^2\right)t_{\mu\nu} = -m^2\,t_{\mu\nu},
\qquad
\left(-\bar\nabla^2 + 2H^2 + m^2\right)t_{\mu\nu} = 0.
\label{eq:TTmassive}
\end{equation}
The shift $6H^2=2(d-1)H^2$ is fixed by requiring that $m^2=0$ reproduces the massless graviton on the Einstein slice. The Higuchi bound is then derived by decomposing the massive spin-2 field in helicity modes. A convenient way to do so is via a Stückelberg decomposition \cite{ArkaniHamed:2002sp},
\begin{equation}
h_{\mu\nu} \to h_{\mu\nu} + \bar\nabla_\mu A_\nu + \bar\nabla_\nu A_\mu
+2\bar\nabla_\mu\bar\nabla_\nu \pi.
\label{eq:Stueck}
\end{equation}
After substituting \cref{eq:Stueck} into \cref{eq:FPaction} and diagonalizing the kinetic terms, one finds that the helicity-0 mode $\pi$ acquires an effective kinetic term proportional to \linebreak
$(m^2-2H^2)$. Up to an overall normalization and total derivatives, the quadratic action for $\pi$ takes the schematic form
\begin{equation}
S_{\pi}^{(2)} \sim \int d^4x \sqrt{-\bar g}\,
\left(m^2-2H^2\right)\,(\bar\nabla \pi)^2,
\label{eq:pikin}
\end{equation}
where the omitted terms include mass terms for $\pi$ and mixing terms. The Higuchi bound \cite{Higuchi:1986py}
\begin{equation}
m^2 \ge 2H^2,
\label{eq:Higuchi}
\end{equation}
is the condition for the helicity-0 mode of a massive graviton to have a non-negative kinetic term. For $m^2<2H^2$ that kinetic term is ghost-like and the theory is non-unitary. At saturation, $m^2=2H^2$, the theory becomes partially massless \cite{Deser:2001pe} and an additional scalar gauge symmetry appears, which removes the helicity-0 mode.

The Higuchi bound is a statement about unitary representations of the de Sitter isometry group, so it is expected to be robust under small deformations. In particular we expect it to hold in quasi-de Sitter or inflationary scenarios \cite{Arkani-Hamed:2015bza,Bordin:2016ruc, Grisa:2009yy, Kolb:2023dzp}. For $\mathrm{dS}$ maxima the Higuchi ghost is present at arbitrarily short wavelength, whereas the instability of a maximum is only relevant at horizon scales.
In an effective description arising from the presence of extra dimensions the mass $m$ of massive gravitons is not a free parameter but is determined by the KK spectrum, and so the bound becomes a constraint on the compactification data. From a group theoretic perspective, the massive modes of the higher-dimensional graviton organize themselves into Fierz-Pauli fields in lower dimension. As the simplest example, the first massive KK mode in a compactification of the form
\begin{eqaed}
 \mathrm{dS}_4\times S^1
\end{eqaed}
must satisfy the bound. In fact, in this geometry the modes must organize in representations of the isometry group 
\begin{eqaed}
    SO(1,4)\times U(1)
\end{eqaed}
where $SO(1,4)$ is the isometry group on $\mathrm{dS}_4$ and $U(1)$ is generated by translations along $S^1$. Modes can be chosen to furnish simultaneous representations, so we can decompose any field into eigenmodes of $U(1)$,
\begin{equation}
-i\partial_y\,\Phi_n(x,y)=m_n\,\Phi_n(x,y),
\end{equation}
with $m_n$ the conserved $U(1)$ charge and $y$ a coordinate on $S^1$. Since $S^1$ is compact, the spectrum of charges is discrete, and we obtain one neutral sector with $m_0=0$ together with an infinite set of charged sectors labeled by $n$. Each charged sector describes modes that carry non-zero momentum along $S^1$ and therefore appear as massive fields from the perspective of $\mathrm{dS}_4$. The Higuchi bound is the statement that there are no unitary massive spin-2 representations of $SO(1,4)$ with $m^2<2H^2$. Since the charges $m_n$ are fixed by the geometry of $S^1$, the bound must in principle constrain the internal geometry.

\section{The Higuchi bound under compactification} \label{sec:iv}

We conjecture that a spectrum of light spin-2 modes with masses violating the Higuchi bound \cite{Higuchi:1986py} does not arise from a consistent compactification of higher-dimensional gravity to $\mathrm{dS}_4$. This has, for example, been shown to hold explicitly for rigid geometries of the form $\mathrm{dS}_4\times X$, with $X$ Einstein and positively curved, supported by a higher-dimensional cosmological constant \cite{Hinterbichler:2013kva}. Let us consider here as starting point the rigid $\mathrm{dS}_4\times S^1$ in static coordinates
\begin{eqaed}
    ds^2= -(1-H^2\,r^2)dt^2+\dfrac{dr^2}{1-H^2\,r^2}+r^2d\Omega^2+dy^2,\,\,\,y\simeq y+L,
\end{eqaed}
supported by some sources.\footnote{Note that this metric interpolates directly to Minkowski ($H\to 0$) as well as to AdS ($H\to iH$).} This metric yields the Einstein tensor
\begin{eqaed}
    G^{\,\,\,\,M}_N=3H^2 \times \mathrm{diag}\left(-1,-1,-1,-1,-2\right).
\end{eqaed}
The energy momentum tensor sourcing the geometry necessarily violates the null energy condition (NEC). In fact, for
\begin{eqaed}
    T^{\,\,\,\,M}_N=\dfrac{3H^2}{8\pi G_5} \times \mathrm{diag}\left(-1,-1,-1,-1,-2\right),
\end{eqaed}
we can take a null vector 
\begin{eqaed}
    k^M=(\frac{1}{\sqrt{1-H^2\, r^2}},0,0,0,1)
\end{eqaed}
and show that
\begin{eqaed}
    G_{MN}k^M k^N=-3H^2<0.
\end{eqaed}
Supporting a $\mathrm{dS}_4$ geometry in the presence of a compact extra direction therefore requires NEC violating sources \cite{Maldacena:2000mw}. This generalizes to $d+p$ dimensions. Consider
\begin{eqaed}
    ds^2= -(1-H^2\,r^2)dt^2+\dfrac{dr^2}{1-H^2\,r^2}+r^2d\Omega^2_{d-2}+g_{ij}(y)dy^idy^j.
\end{eqaed}
Here the curvature decomposes as 
\begin{eqaed}
    R_{\mu\nu}= R_{\mu\nu}^{ext}= (d-1) H^2 g_{\mu\nu}\,,\quad R_{ij}= R_{ij}^{int}(y)\,,
\end{eqaed}
\begin{eqaed}
    R= R^{ext}+R^{int}= d(d-1) H^2+R^{int}(y).
\end{eqaed}
Let us consider a null vector $k^M$ such that
\begin{eqaed}
g_{MN}k^M k^N=g_{\mu\nu}k^\mu k^\nu+g_{ij}k^i k^j=0.
\end{eqaed}
We then have
\begin{eqaed}
 G_{MN}k^M k^N=(R_{ij}^{int}(y)-(d-1)\,H^2\,g_{i j}) k^i k^j.
\end{eqaed}
For a Ricci flat manifold we have
\begin{eqaed}
 G_{MN}k^M k^N=-(d-1)\,H^2\,g_{i j} k^i k^j<0,
\end{eqaed}
since the internal manifold is Euclidean and positive definite. For curved manifolds we take $k^i$ along the direction of smallest internal Ricci curvature, for which
\begin{eqaed}
    R^{int}_{ij}k^i k^j\leq  \dfrac{R^{int}}{p}\,|k^i|^2,
\end{eqaed}
since the smallest eigenvalue of a symmetric tensor is at most its trace divided by the dimension.
For any negatively curved internal manifold, $R^{int}(y)<0$, one must then violate the null energy condition in order to support compactifications to de Sitter. Large enough positive curvature avoids this, but positive internal curvature contributes negatively to the 4d vacuum energy, so it works against building a $\mathrm{dS}$ extremum in the first place.
No classical source violates the NEC, so we explore the possibility of realizing this with either Casimir energies or negative tension extended objects (morally $Op$-planes). Note that AdS requires no such sources, and by triviality neither does Minkowski, which is well known in the context of string theory model building (see e.g. \cite{Gibbons:1984kp,Grana:2005jc,Blumenhagen:2013fgp,Basile:2020mpt, VanRiet:2023pnx}).

Our proposal is similar in spirit to \cite{Lust:2019lmq}, where it was argued that since the higher spin excitations of the string necessarily violate the Higuchi bound, there is no valid effective IR description for these modes. The string excitations corresponding to those higher spin modes have lengths $\ell_{str}>1/H$, so they are not particle-like within a local patch, and one must resort to a UV description. If $m_{\rm KK}<H$ then the internal dimensions are larger than the horizon, $L>1/H$, and similarly we expect UV physics to enter here.
\subsection{Compactification on \texorpdfstring{$S^1$}{S1} with Casimir energy}

A simple class of examples is obtained by compactifying on a circle of radius $R$ and generating a positive four dimensional vacuum energy via the one-loop Casimir energy. We start with an unwarped metric ansatz
\begin{eqaed}
    ds^2= e^{2\alpha\phi(x)}g_{\mu\nu}dx^{\nu}dx^{\mu}+e^{2\beta\phi(x)} dy^2,
\end{eqaed}
with
\begin{eqaed}
    \alpha=\frac{1}{\sqrt{(d-1)(d-2)}}=\frac{1}{\sqrt{6}},\qquad \beta=-(d-2)\alpha=-\frac{2}{\sqrt{6}},
\end{eqaed} 
fixed by the requirement that the resulting 4d theory is Einstein-Hilbert gravity with a canonically normalized scalar. Before introducing sources, consider the energy momentum tensor required to support de Sitter slices on $g_{\mu\nu}$,
\begin{align}
G^{\mu}{}_{\nu}{}&=e^{-2\phi/\sqrt{6}}\left(-3H^2 \delta^\mu_\nu-\partial^{\mu}\phi\,\partial_{\nu}\phi+\frac{1}{2}\,\delta^{\mu}{}_{\nu}(\partial\phi)^2\right), \\
G^{5}{}_{5}&=e^{-2\phi/\sqrt{6}}\left(-6H^2+\frac{\sqrt{6}}{2}\,g^{\mu\nu}\partial_\mu\partial_\nu\phi+\frac{1}{2}(\partial\phi)^2\right),
\end{align}
with mixed components zero. The requirement of having a rigid $\mathrm{dS}_4$ is then equivalent to demanding that the radion $\phi$ is \textit{classically} stabilized at some $\phi_\star$, such that
\begin{eqaed}
    \partial_\mu\phi\big\vert_{\phi\to\phi_{\star}}=0.
\end{eqaed}
By classically we mean that we treat the radion as stabilized even if it sits at a maximum, that is, in a tachyonic direction.
In this case the spin-2 perturbations around the background 4d metric satisfy the eigenvalue equation
\begin{equation}
\left(\Delta_L - m_n^2\right)h^{(n)}_{\mu\nu} = 0,
\label{eq:4dmodeeq2}
\end{equation}
with $\Delta_L$ the Lichnerowicz operator of the background 4d metric. In the absence of warping, the only dependence on $y$ enters through the periodicity conditions
\begin{eqaed}
    y\simeq y+2\pi R_0,
\end{eqaed}
so that the KK gravitons gain a mass
\begin{eqaed}
    m_n^2=\dfrac{n^2}{R_0^2}e^{-2(\beta-\alpha)\phi}=\dfrac{n^2}{R^2},
\end{eqaed}
where we have written it in the end in terms of the physical length of the interval. 

The Higuchi bound in its original formulation assumes just a massive graviton on a fixed background, so we keep the number of additional ingredients as small as possible. At tree level we assume a positive 5d cosmological constant, supplemented by the one-loop contribution of the 5d graviton running around the circle, the Casimir energy. This is not an extra assumption as Casimir energy is generic in non-supersymmetric theories on compact backgrounds \cite{Appelquist:1983vs, ArkaniHamed:2007gg}.
The 5d action is
\begin{equation}
S_5=\frac{M_5^{3}}{2}\int d^5x\,\sqrt{-G}\left[R_5-\Lambda_5\right],
\end{equation}
and after compactification on $S^1$ we obtain
\begin{equation}
S_4=\frac{M_{\rm Pl}^{2}}{2}\int d^4x\,\sqrt{-g}\left[R-(\partial\phi)^2-V(\phi)\right].
\end{equation}
The two contributions to the 4d potential are
\begin{equation}
V(\phi)=V_{\Lambda}+V_{\mathrm{Casimir}}=\Lambda_5 e^{\sqrt{\frac{2}{3}}\phi}-\frac{1}{M_{\rm Pl}^{2}}\frac{15\zeta(5)}{32\pi^6}\frac{1}{R_0^4}e^{4\sqrt{\frac{3}{2}}\phi},
\end{equation}
where the factor of 15 refers to the degrees of freedom of the 5d graviton. Since the bosonic contribution dominates, the Casimir term is negative, and as anticipated above this is what allows us to construct a $\mathrm{dS}$ solution.
Extremizing the potential gives a de Sitter maximum at
\begin{eqaed}
\phi_{\star}= \frac{\sqrt{6}}{10}\log \left(\frac{16\, \pi ^6 \Lambda_5\, M_{\rm Pl}^2\, R_0^4}{45\, \zeta(5)}\right),
\end{eqaed}
so that the potential takes the form
\begin{eqaed}
V(\phi)\big\vert_{\phi=\phi_{\star}}=\frac{5}{6}\,\Lambda_{5}\left(\frac{16\,\pi^{6}\,M_{\rm Pl}^{2}\, R_0^{4}\, \Lambda_{5}}{45\,\zeta(5)}\right)^{1/5}.
\end{eqaed}
This takes a simpler form in terms of the physical length of the circle, $R=R_0 e^{-\sqrt{3/2}\phi}$,
\begin{eqaed}
V(\phi)\big\vert_{\phi=\phi_{\star}}=\frac{75\,\zeta(5)}{32\,M_{\rm Pl}^{2}\,\pi^{6}\,R^{4}}.
\end{eqaed}
More generally, balancing Casimir against a potential of the type
\begin{eqaed}
    V= \lambda R^{-n}
\end{eqaed}
leads to a maximum with
\begin{eqaed}
V(\phi)\big\vert_{\phi=\phi_{\star}}=\left(\frac{4}{n}-1\right)\frac{15\,\zeta(5)}{32\,M_{\rm Pl}^{2}\,\pi^{6}\,R^{4}}.
\end{eqaed}
Demanding a positive cosmological constant then requires $0<n<4$. The case above is $n=2/3$, the five dimensional cosmological constant scaling as $R^{-2/3}$ once written in terms of the physical radius.

This setup realizes the dark dimension \cite{Montero:2022prj}, as one expects from balancing Casimir energy. The Hubble scale is given by
\begin{eqaed}
H=\sqrt{V(\phi)/6}=\frac{5\,\sqrt{\zeta(5)}}{8\,M_{\rm Pl}\,\pi^{3}\,R^{2}}.
\end{eqaed}
Demanding that the physical size of the compact direction is larger than the 5d Planck length
\begin{eqaed}
    2 \pi R> M_5^{-1}
\end{eqaed}
imposes
\begin{eqaed}
    M_{\rm Pl} \,2\pi R= (M_5\, 2\pi R)^{3/2}> 1.
\end{eqaed}
Then, the Hubble scale and the size of the circle are related as
\begin{eqaed}
    H\,R=\frac{5\,\sqrt{\zeta(5)}}{8\,\,\pi^{3}\,}\dfrac{\,2\pi}{M_{\rm Pl}\,2\pi R}< 1,
\end{eqaed}
and including numerical factors,
\begin{eqaed}
    H\,R\simeq 0.13\,\dfrac{1}{(M_{5}\,2\pi R)^{3/2}}.
\end{eqaed}
The mass of the first massive graviton is $R^{-1}$, so the Higuchi bound proper reads $HR<1/\sqrt{2}$, and the setup with the minimal set of 5d ingredients in fact satisfies the stronger condition $HR\lesssim 0.13$. Approaching saturation is equivalent to shrinking the circle down to the species scale. In this sense one cannot, in pure gravity, start from a consistent 5d theory on a circle and break the Higuchi bound using the standard ingredients for violating the null energy condition. The same follows immediately for an internal Ricci-flat manifold of dimension $p$, provided the leading cycle is the volume, that is, provided no sub-cycle grows large.

\subsection{Warped \texorpdfstring{$\mathrm{dS}_4\times S^1/\mathbb{Z}_2$}{dS4xS1/Z2} with negative tension at the boundaries}

A second example is a strongly warped $S^1/\mathbb{Z}_2$ interval, where one can engineer a large hierarchy between the Hubble length and the coordinate size of the extra dimension. Stabilizing the interval modulus is non-trivial in RS-like setups \cite{Randall:1999ee, Randall:1999vf}. There are proposals both with \cite{Luty:1999cz} and without \cite{Goldberger:1999uk} supersymmetry, quite reminiscent of stabilizing moduli with fluxes in string theory. We will instead take the manifold as rigid, assuming that stabilization has taken place and that its effects on the resulting dynamics are subdominant. Consider static coordinates with a warped compactification ansatz to 4d de Sitter
\begin{equation}
ds^{2}=A(y)^{2}\left[-(1-H^{2}r^{2})\,dt^{2}+\frac{dr^{2}}{1-H^{2}r^{2}}+r^{2}d\theta^{2}+r^{2}\sin^{2}\theta\,d\varphi^{2}\right]+dy^{2}.
\end{equation}
This has Einstein tensor
\begin{eqaed}
    G_\mu^\nu=\delta_\mu^\nu \frac{3 \left(-H^2+ A(y) A''(y)+A'(y)^2\right)}{A(y)^2},\quad G^5_5= \frac{6 \left(-H^2+A'(y)^2\right)}{A(y)^2}.
\end{eqaed}
Requiring a background locally of the form $\mathrm{dS}_4\times S^1$ fixes  
\begin{eqaed}
    A''(y)=0,\qquad A(y)=1+\lambda\,y,
\end{eqaed}
so that the metric is
\begin{equation}
ds^{2}=(1+\lambda y)^{2}\left[-(1-H^{2}r^{2})\,dt^{2}+\frac{dr^{2}}{1-H^{2}r^{2}}+r^{2}d\theta^{2}+r^{2}\sin^{2}\theta\,d\varphi^{2}\right]+dy^{2},
\end{equation}
with Einstein tensor
\begin{eqaed}
    G_\mu^\nu=\delta_\mu^\nu \frac{3 \left(-H^2+\lambda^2\right)}{(1+\lambda\,y)^2},\quad G^5_5= \frac{6 \left(-H^2+\lambda^2\right)}{(1+\lambda\,y)^2}.
\end{eqaed}
This requires additional bulk sources to support the geometry, unless $\lambda=H$, in which case the energy momentum tensor vanishes exactly.

This is a known solution \cite{Kaloper:1999sm, Karch:2000ct} to an interval with pure tension codimension-1 sources at the boundaries. We recall the induced energy momentum tensor at the $y=0,L$ boundaries\footnote{This form of the energy-momentum tensor arises due to the $\mathbb{Z}_2$ Israel boundary conditions \cite{Israel:1966rt} at the ends of the interval.} 
\begin{eqaed}
    S_{\mu\nu}=-2M_5^3\left(K_{\mu\nu}-K\,\tilde g_{\mu\nu}\right),
\end{eqaed}
with the metric on the slices
\begin{eqaed}
    \tilde g_{\mu\nu}(x,y)=A(y)^2 g_{\mu\nu},
\end{eqaed}
and the extrinsic curvature
\begin{eqaed}
    K_{\mu\nu}=\pm\frac{1}{2}\partial_y(A(y)^2)\, g_{\mu\nu}(x)=\pm\frac{A'}{A}\tilde g_{\mu\nu},
\end{eqaed}
with the sign depending on the normal vector to the boundary.
The codimension-1 sources at the boundaries must have energy momentum tensor
\begin{eqaed}
    S^{0}_{\mu\nu}=6M_5^3H \tilde g_{\mu\nu}(x,0),\quad S^{L}_{\mu\nu}=-6M_5^3\dfrac{H}{1+HL} \tilde g_{\mu\nu}(x,L),
\end{eqaed}
such that they are pure tension sources with tensions
\begin{eqaed}
    \lambda^{0}=6M_5^3H,\quad \lambda^{L}=-6M_5^3\dfrac{H}{1+HL}.
\end{eqaed}
A NEC violating source is then unavoidable, as one of the branes must have negative tension. This is not too unexpected, given that the interval with $\mathbb{Z}_2$ boundary conditions is a $\mathbb{Z}_2$ orbifold of the circle.

We now take the setup as given and focus on the bulk geometry. One can take a Minkowski limit by setting
\begin{eqaed}
    \lambda,H\to 0,
\end{eqaed}
but analytically extending to an $\mathrm{AdS}_4\times S^1$ geometry is not possible without bulk sources, as a bulk vacuum solution would be obtained by taking
\begin{eqaed}
    \lambda,H\to iH\,,
\end{eqaed}
such that the metric would be complex.
We then compute the KK spectrum for gravitons around this background, leading to an equation for the transverse-traceless graviton modes (see Appendix \ref{sec:kkwarp} for details on the computation)
\begin{eqaed}
    (\Delta_L-6H^2)h_{\mu\nu}-A(y)^2 h_{\mu\nu}''-4A(y)A'(y)h_{\mu\nu}'=0.
\end{eqaed}
After imposing separation of variables
\begin{eqaed}
    h_{\mu\nu}=t_{\mu\nu}(x)\psi(y)
\end{eqaed}
and identifying the Fierz-Pauli mass from the Lichnerowicz operator as
\begin{eqaed}
    (\Delta_L-6H^2)t_{\mu\nu}=-m^2\,t_{\mu\nu},
\end{eqaed}
we obtain
\begin{eqaed}
\frac{m^2\,\psi (y)}{A(y)^2}+\frac{4 A'(y)\,\psi '(y)}{A(y)}+\psi ''(y)=0.
\end{eqaed}
This Sturm-Liouville system has an exact solution for $A(y)=1+Hy$,
\begin{equation}
\psi(y)=\psi_0\left[(1+Hy)^{-\delta_+}-\frac{\delta_+}{\delta_-}(1+Hy)^{-\delta_-}\right],
\end{equation}
with 
\begin{equation}
\delta_\pm=\frac{3}{2}\pm \sqrt{\frac{9}{4}-\frac{m^2}{H^2}}.
\end{equation}
This then leads to a discrete mass spectrum after imposing Neumann boundary conditions on the modes of the graviton \cite{Randall:1999vf}
\begin{eqaed}
    h_{\mu\nu}'\big\vert_{y=0,L}=0.
\end{eqaed}
The eigenvalue equation then gives
\begin{eqaed}
    m_n^2= \dfrac{9}{4}H^2+n^2m_{\rm KK}^2,
\end{eqaed}
with 
\begin{eqaed}
    m_{\rm KK}^2=\dfrac{\pi^2}{\ell_{conf}^{\,2}},\quad    \ell_{conf}=\int_0^L \dfrac{1}{A(y)}dy=\dfrac{H}{\log{(1+H L)}}.
\end{eqaed}
We see that this setup never violates the Higuchi bound, as warping gaps the tower
\begin{eqaed}
    m_n^2\geq \dfrac{9}{4}H^2>2H^2,
\end{eqaed}
so the Higuchi bound is always satisfied, independent of $H$ and $L$.
The same background was studied in \cite{Antoniadis:2015txa}, where it was argued that the KK gravitons did not satisfy the Higuchi bound because they only saw the raw Laplacian in 4d rather than the full Lichnerowicz operator, and so led to no violation of unitarity in 4d. Here we find instead that they cannot violate unitarity in 4d for a simpler reason, namely that their masses always sit above Hubble. There are two relevant limits. For $H\,L\ll 1$ we recover the standard unwarped case,
\begin{eqaed}
    m_n^2\simeq \dfrac{\pi^2 n^2}{L^2}\gg H^2,
\end{eqaed}
while in the strongly warped regime, $H\,L\gg1$, the KK masses organize themselves so that the bound is also never violated,
\begin{eqaed}
    m_n^2\simeq\dfrac{9}{4}H^2+ \dfrac{\pi^2 n^2 H^2}{\log(HL)^2}>2 H^2.
\end{eqaed}
In this limit the species scale no longer matches the 5d Planck mass, which may mean that the standard species counting  is not valid outside the strict Minkowski limit \cite{Calderon-Infante:2023ler,Raucci:2026fzp}. The same restriction applies to the emergent string conjecture \cite{Lee:2019wij}, which is only formulated for decompactifications to Minkowski. Indeed, for $H\to 0$ we recover Minkowski, the solution collapses to the unwarped case and $\Lambda_{sp}=M_5$. Extending the warped solution to AdS \cite{Karch:2000ct}, $H\to-i/\ell_{\mathrm{AdS}}$, only makes sense in the weakly warped limit, where the KK gravitons have masses
\begin{eqaed}
m_n^2\simeq-\dfrac{9}{4}\dfrac{1}{\ell_{\mathrm{AdS}}^2}+ \dfrac{\pi^2 n^2}{L^2}.
\end{eqaed}
Repeating the computation for the KK masses of bulk scalars, one finds that they always respect the BF bound \cite{Breitenlohner:1982bm, Metsaev:2003cu}
\begin{eqaed}
    m^2>-\dfrac{9}{4}\dfrac{1}{\ell_{\mathrm{AdS}}^2},
\end{eqaed}
while the KK gravitons may not satisfy it, as the bound is modified for particles with non-zero spin. For spin-2 the bound requires that there are no tachyonic modes \cite{Metsaev:2003cu}, so that 
\begin{eqaed}
    m^2>0.
\end{eqaed}
Positive masses then require $L<2\pi\ell_{\mathrm{AdS}}/3$, again the weakly warped regime, so the analytically continued AdS satisfies the BF bound and is free of tachyonic instabilities.
Note that one can imitate the dark dimension in this setup, since the Hubble scale and the length of the interval are decoupled. 

Momentarily restoring Planck masses, the cosmological constant is 
\begin{eqaed}
    \Lambda_{cc}=M_{\rm Pl}^2 H^2,
\end{eqaed}
and reproducing the dark dimension scenario requires 
\begin{eqaed}
    \dfrac{1}{\ell_{conf}^4}\simeq \Lambda_{cc},
\end{eqaed}
which forces the weakly warped regime\footnote{Dark dimension-like constructions in five dimensional Hořava-Witten theory have been argued to also exclude strong warping \cite{Reig:2025gch,Blumenhagen:2026rgu}, although for a different phenomenological reason.} and
\begin{eqaed}
    L\simeq\dfrac{\ell_{Pl}}{\sqrt{H/M_{\rm Pl}}}.
\end{eqaed}
The scale of the extra dimension is then $L=\Lambda_{cc}^{-1/4}$, as in the dark dimension. In this setup this does not however arise as a fundamental relation between the size of the extra dimensions and the vacuum energy, but has been imposed by hand.

\section{The KK spectrum on group manifolds}
 
In the previous section the ingredients used to evade the null energy condition, Casimir energy and negative tension branes, turned out to respect the Higuchi bound on their own. However, internal curvature can also produce NEC violating contributions. The internal curvature acts as a source in the lower dimensional equations of motion, so that internal manifolds with negative scalar curvature contribute positive vacuum energy. In this section we argue that the bound forbids a de Sitter extremum whose curvature contribution is parametrically small, and we then test this against explicit group manifolds. Possible issues with using curvature in cosmological solutions were also studied in \cite{Andriot:2025cyi, Coudarchet:2023mfs}, where it was argued to work against scale separation. We focus on the simpler case of group manifolds, where the curvature is fixed algebraically by the structure constants of the underlying algebra and the KK spectrum by its representation theory. Consider a product space $\mathcal{N}_{d}\times\mathcal{M}_p$, with metric
\begin{eqaed}\label{eq:product-ansatz}
    ds^2_D = e^{2\alpha\phi(x)}\,\tilde{g}_{\mu\nu}(x)\,dx^\mu dx^\nu + e^{2\beta\phi(x)}\,g_{mn}(y)\,dy^m dy^n\,, \qquad D = d + p\,,
\end{eqaed}
with $p$ compact dimensions, where $\phi$ is the volume modulus and the constants
\begin{eqaed}
    \alpha^2 = \frac{p}{2(d-2)(d+p-2)}\,, \qquad \beta = -\,\frac{d-2}{p}\,\alpha
    \label{eq:alphabeta}
\end{eqaed}
are chosen so that the $d$-dimensional theory is in Einstein frame with a canonically normalized scalar. After compactification we then obtain an action of the form
\begin{eqaed}\label{eq:reduced-action}
    S_d = \frac{1}{2\kappa_d^2}\int d^dx\,\sqrt{-\tilde{g}}\,\left[\tilde{R} - \frac{1}{2}(\partial\phi)^2 - V(\phi)\right], \qquad V(\phi) = -\,R_{\mathcal{M}_p}\,e^{2(\alpha-\beta)\phi}\,,
\end{eqaed}
where $R_{\mathcal{M}_p}$ is the scalar curvature of the internal metric $g_{mn}$ and $\kappa_d^{-2} = \mathrm{Vol}(\mathcal{M}_p)\,\kappa_D^{-2}$. We set $\kappa_d=1$ throughout. As anticipated, a negatively curved internal space contributes positively to the potential. We recall the scaling of the mass of KK modes in an isotropic decompactification limit,
\begin{eqaed}\label{eq:kk-scaling}
    m_{\rm KK}^2 \sim \frac{e^{2(\alpha-\beta)\phi}}{L^2}\,,
\end{eqaed}
with $L$ the length scale of the internal metric. The curvature scales as
\begin{eqaed}
   R_{\mathcal{M}_p}\propto m_{\rm KK}^2\,.
\end{eqaed}
Both carry the same power of the volume modulus, so their ratio is a pure number fixed by the internal geometry. Using curvature as an ingredient for $\mathrm{dS}$ therefore sets $H^2\sim m_{\rm KK}^2$, and one cannot guarantee the Higuchi bound is parametrically satisfied. For example, with only two terms in the  potential \cite{Andriot:2025cyi},
\begin{eqaed}\label{eq:two-term-potential}
    V(\phi) = -\,R_{\mathcal{M}_p}\,e^{2(\alpha-\beta)\phi} - V_1\,e^{-2\gamma\phi}\,, \qquad R_{\mathcal{M}_p}<0\,.
\end{eqaed}
Assuming $V_1,\,\gamma>0$ there will be a $\mathrm{dS}$ extremum given by
\begin{eqaed}\label{eq:ds-extremum}
V(\phi_*) = |R_{\mathcal{M}_{p}}|\,e^{2(\alpha-\beta)\phi_*}\left(1+\frac{(\alpha-\beta)}{\gamma}\right) \propto m_{\rm KK}^2\,,
\end{eqaed}
so that $H^2 = \frac{V(\phi_*)}{(d-1)(d-2)}\simeq |V_R|$. One way to avoid this is to consider tuning the potential to have vanishing vacuum energy and adding a small third contribution on top, so that the vacuum energy is parametrically below $|V_R|$. A de Sitter extremum on a negatively curved internal manifold cannot then have vacuum energy parametrically above its curvature, that is, it must satisfy $V\lesssim |V_R|$.
 
Let us consider group manifolds, which have been used in explicit classical de Sitter constructions \cite{Andriot:2020vlg} (for a review and their context within string theory model building see \cite{Andriot:2026lac}). Here we go briefly over the most relevant aspects. Consider a metric $g_{mn}$ with coordinates $y^m$ and a local left-invariant vielbein basis
\begin{eqaed}
    e^a_m\,, \qquad g_{mn} = \delta_{ab}\, e^a_m e^b_n\,,\qquad   e^a = e^a_m\, dy^m\,, \qquad \partial_a = e^m_a\,\partial_m\,,
\end{eqaed}
as well as a spin connection 
\begin{eqaed}
    \left[\partial_a, \partial_b\right] = \left(w^{\,c}_{\,\,\,\,a\,b} - w^{\,c}_{\,\,\,\,b\,a}\right)\partial_c,\qquad     de^a + w^{\,a}_{\,\,\,\,b\,c}\, e^b \wedge e^c = 0\,,\quad    f^{\,a}_{\,\,\,\,b\,c} \equiv w^{\,a}_{\,\,\,\,b\,c} - w^{\,a}_{\,\,\,\,c\,b}\,.
\end{eqaed}
A group manifold is such that the $f^{\,a}_{\,\,\,\,b\,c}$ are constant and correspond to the structure constants of a Lie algebra. The basis vectors then furnish a basis of generators of the algebra
\begin{eqaed}
    \left[\partial_a, \partial_b\right] = f^{\,c}_{\,\,\,\,a\,b}\,\partial_c\,,
\end{eqaed}
and in particular the curvature depends only on the structure constants, as the metric has constant coefficients in the left invariant basis. For a left invariant metric on a  unimodular algebra one finds \cite{Milnor:1976}
\begin{eqaed}\label{eq:curvature-structure-constants}
    R = -\,\frac{1}{2}\,\delta^{cd}\, f^{\,a}_{\,\,\,\,b\,c}\, f^{\,b}_{\,\,\,\,a\,d}
        \,-\, \frac{1}{4}\,\delta_{ad}\,\delta^{be}\,\delta^{cf}\, f^{\,a}_{\,\,\,\,b\,c}\, f^{\,d}_{\,\,\,\,e\,f}\,.
\end{eqaed}
The internal manifold is then itself a group, $\mathcal{M}_p \simeq G$, when its algebra $\mathfrak{g}$ is a compact Lie algebra. For physical reasons motivated in string theory, we restrict our analysis to manifolds of dimension at most seven. Looking at the classification of compact Lie algebras \cite{Helgason}, i.e.\ algebras with completely antisymmetric structure constants $f_{abc} = f_{[abc]}$, we see that for dimensions $p \leq 7$ they are all of the form $\mathfrak{g} = \mathfrak{u}(1)^{q} \oplus \mathfrak{su}(2)^{k}$, with $q + 3k = p$. More generally, we can consider a non-compact Lie group $G$ and a discrete subgroup $\Gamma$ such that $\mathcal{M}_p \simeq G/\Gamma$ is compact\footnote{The underlying algebra must be unimodular, $f^{\,a}_{\,\,\,\,a\,b} = 0$, in order to admit a compact quotient, but this is in general not a sufficient condition. For nilpotent algebras a lattice exists if and only if the algebra admits a basis with rational structure constants \cite{Malcev}.}. Among these quotients are the simplest examples of negatively curved manifolds. Fixing the algebra does not fix the overall volume modulus, so we consider a manifold of the form
 \begin{eqaed}
    ds^2= e^{2\alpha\phi(x)}g_{\mu\nu}dx^{\nu}dx^{\mu}+L^2e^{2\beta\phi(x)} g_{mn}dy^m dy^n,
\end{eqaed}
with $g_{mn}$ the metric on a group manifold and $\alpha$, $\beta$ as in \cref{eq:alphabeta}. Given the curvature $R[g_{mn}]$, the term entering the $d$-dimensional equations of motion is
\begin{eqaed}
   V_R= -\dfrac{e^{2(\alpha-\beta)\phi}}{L^2}R[g_{mn}].
\end{eqaed}
Finding the spectrum of spin-2 KK modes requires solving the Laplace equation on $g_{ij}$, and the resulting masses pick up the same overall scaling
\begin{eqaed}
 m_{\rm KK}^2\propto  \dfrac{e^{2(\alpha-\beta)\phi}}{L^2}.
\end{eqaed}
Then, if the group manifold admits arbitrarily small KK masses relative to the curvature it would violate the Higuchi bound. 

As a warmup example take $\mathcal{N}_d\times S^3$, with $S^3\simeq SU(2)$ and $f^{\,a}_{\,\,\,\,b\,c}=\epsilon_{abc}$, for which \cref{eq:curvature-structure-constants} gives $R[g_{mn}]=3/2$. For a bi-invariant metric the harmonics of the Laplacian organize themselves into representations $(j,j)$ of the isometry group $SO(4)=(SU(2)_L\times SU(2)_R)/\mathbb{Z}_2$,
\begin{eqaed}\label{eq:s3-spectrum}
    \tilde m^2_j=j(j+1)\,,\qquad d_j=(2j+1)^2\,,\qquad j=0,\tfrac{1}{2},1,\dots
\end{eqaed}
The ratio between the lightest KK mode $j=1/2$ and the curvature is then fixed
\begin{eqaed}\label{eq:s3-ratio}
    \dfrac{m^2_{\rm KK}}{|V_R|}=\dfrac{\tilde m^2_1}{|R|}=\dfrac{1}{2}.
\end{eqaed} 
We compute the spectrum and the Casimir energy for a relatively simple negatively curved manifold, $\rm Nil_3$ \cite{Scott:1983}, with metric
\begin{eqaed}
    ds^2=dx^2 + dy^2 +(dz-x dy)^2,
\end{eqaed}
so that $d=4$, $p=3$.
In order to have a compact manifold we consider a lattice $\Gamma=\left<\gamma_1,\gamma_2,\gamma_3\right>$, with 
\begin{eqaed}
    \gamma_1=(\ell_x,0,0),\quad    \gamma_2=(0,\ell_y,0),\quad    \gamma_3=(0,0,\ell_z).
\end{eqaed}
For $\Gamma$ to be a subgroup of $\rm Nil_3$ it must be closed under the group action
\begin{eqaed}
    (x,y,z).(x',y',z')=(x+x', y+y', z+z' +x y'),
\end{eqaed}
and since 
\begin{eqaed}
    \gamma_1\,\gamma_2\,\gamma_1^{-1}\,\gamma_2^{-1}=(0,0,\ell_x \ell_y),
\end{eqaed}
this requires
\begin{eqaed}
    \ell_z=\dfrac{\ell_x \ell_y}{N}
\end{eqaed}
with $N$ some integer. Here $\ell_x$ and $\ell_y$ label smooth deformations of the compact $\rm Nil_3$, whereas $N$ labels topologically distinct choices of lattice. The scalar Laplacian is
\begin{eqaed}
\nabla^2f=(1+x^2)\frac{\partial^2 f}{\partial z^2}
+2x\frac{\partial^2 f}{\partial y\,\partial z}
+\frac{\partial^2 f}{\partial y^2}
+\frac{\partial^2 f}{\partial x^2}.
\end{eqaed}
Since $\partial_y$ and $\partial_z$ are commuting Killing vectors on $\rm Nil_3$ we have
\begin{eqaed}
    f= h(x)\, e^{i\lambda_y y}\,e^{i\lambda_z z},
\end{eqaed}
with $\lambda_y=\frac{2\pi j}{\ell_y}$, $\lambda_z=\frac{2\pi N k}{\ell_x\,\ell_y}$, with $j$ and $k$ integer.
 
The eigenvalue equation for $k>0$ is then
\begin{eqaed}
    \frac{h''(x)}{h(x)}-\lambda_y^2-\lambda_z^2-\lambda_z^2 x^2-2 \lambda_y\,\lambda_z\, x=-m^2,
\end{eqaed}
which has a normalizable solution
\begin{eqaed}
    h(x)=c_0 D_{\frac{m^2-\lambda_z^2-\lambda_z}{2 \lambda_z}}\left(x \sqrt{\lambda_z} \sqrt{2}+\frac{\lambda_y\sqrt{2}}{\sqrt{\lambda_z}}\right),
\end{eqaed}
with $D_\nu$ the parabolic cylinder function. Normalizability as $|x|\to\infty$ requires a non-negative integer coefficient, imposing the quantization condition on the mass
\begin{eqaed}
    \frac{m^2-\lambda_z^2-\lambda_z}{2 \lambda_z}=n,\quad m^2_{n,k}=\dfrac{2 N \pi}{\ell_x \ell_y}\left(k (2n+1)+\frac{2 \pi  k^2 N}{\ell_x \ell_y}\right).
\end{eqaed}
Requiring that the solution is well defined on the lattice
\begin{eqaed}
    h(x+\ell_x)\,e^{i\lambda_y y}=h(x)\,e^{i(\lambda_y-\frac{2\pi kN}{\ell_y}) y}
\end{eqaed}
implies $j$ is only defined mod $k N$, and as such the per level degeneracy is $d_{n,k}=k N$.
The solution has a second branch, which becomes evident once one recalls that $\rm Nil_3$ can be understood as a twisted $S^1$ fiber over a $T^2$ base. For $k=0$ the eigenvalue equation becomes\footnote{The complete Laplacian spectrum on $\rm Nil_3$ was obtained in \cite{Andriot:2018tmb}.}
\begin{eqaed}
    m^2_{n,j}=\left(\dfrac{2\pi n}{\ell_x}\right)^2+\left(\dfrac{2\pi j}{\ell_y}\right)^2 .
\end{eqaed}
We focus on the twisted part of the spectrum, as the contribution from the untwisted modes can be made subdominant by tuning $N$. The contribution of the Casimir energy to the scalar potential takes the form \cite{Vassilevich:2003xt}
\begin{equation}
  V \,=\, -\frac{1}{2\,(4\pi)^{d/2}} \int_0^\infty \mathrm{d}t\,
  t^{-1-\frac{d}{2}}\, K(t),
  \qquad
  K(t) \,=\, \sum_{n,k} d_{n,k}\, e^{-t m^2_{n,k}} .
\end{equation}
This double sum can be evaluated in closed form in the limit $\ell_x \ell_y\gg1$ (see Appendix \ref{sec:nil3casimir} for details). For $d=4$ we have
\begin{eqaed}
     V_{d=4} \,=\, -\frac{\zeta(3)}{1280\,\pi^2}\dfrac{N^3}{(\ell_x \ell_y)^2}\,\simeq\,-9.5\times 10^{-5}\dfrac{N^3}{(\ell_x \ell_y)^2}.
\end{eqaed}
We restore dependence on $L$ and $\phi$, and combining this with the internal curvature we have
\begin{eqaed}
    V=   -\dfrac{e^{2(\alpha-\beta)\phi}}{L^2}R_{\rm Nil_3}-\frac{\zeta(3)}{1280\,\pi^2}\dfrac{N^3}{(\ell_x \ell_y)^2}\dfrac{e^{4(\alpha-\beta)\phi}}{L^4}.
\end{eqaed}
The $\mathrm{dS}$ maximum then gives 
\begin{eqaed}
    V_{\mathrm{dS}}=-\dfrac{1}{2 L_{phys}^2}R_{\rm Nil_3},
\end{eqaed}
with $R_{\rm Nil_3}=-\frac{1}{2}$. The lightest KK mass is
\begin{eqaed}
    m^2=\dfrac{N}{L_{phys}^2}\dfrac{2 \pi}{\ell_x \ell_y}\left(1+\frac{2 \pi }{\ell_x \ell_y}\right).
\end{eqaed}
Both $V_{\mathrm{dS}}$ and $m^2$ carry the same power of $L_{phys}$, so the volume modulus drops out of their ratio and the comparison is controlled entirely by the lattice. Choosing $\ell_x, \ell_y\gg 1$ then pushes the lightest KK mass arbitrarily far below the Hubble scale, in apparent violation of the Higuchi bound. That violation signals an inconsistency in the setup, which one should then be able to identify independently. The parameters $\ell_x$ and $\ell_y$ are moduli of $\rm Nil_3$, and a consistent 4d theory requires them to be stabilized. Since they parameterize the $T^2$ base of $\rm Nil_3$, they appear in the compactified action as
\begin{eqaed}
    L\supset \dfrac{\left(\partial \mathcal{V}_{T^2}\right)^2}{\mathcal{V}_{T^2}^2}+\frac{1}{2}\dfrac{\left(\partial U\right)^2}{U^2},
\end{eqaed}
with $\mathcal{V}_{T^2}=\ell_x \ell_y$ and $U=\ell_x/\ell_y$ the volume and complex structure of the $T^2$. The Casimir energy from the twisted modes depends only on $\mathcal{V}_{T^2}$, so $U$ is left unlifted. The untwisted modes can lift $U$, but their contribution is subdominant in the limit $\ell_x,\ell_y\gg 1$. One can consider a more rigid lattice, for which the base becomes $T^2/\mathbb{Z}_4$ and the complex structure is projected out by $\sigma:(x,y)\simeq (-y,x)$ \cite{Grana:2013ila}. This removes $U$, but the volume modulus remains. The lattice fixes the internal volume to $\mathrm{Vol}=\ell_x\ell_y\ell_z=\mathcal{V}_{T^2}^2/N$, so $\mathcal{V}_{T^2}$ enters the four dimensional Planck mass as well, and the potential is modified as
\begin{eqaed}
    V\simeq \dfrac{N}{2\,\mathcal{V}_{T^2}^2}\dfrac{e^{2(\alpha-\beta)\phi}}{L^2}
    -\dfrac{\zeta(3)}{1280\,\pi^2}\dfrac{N^5}{\mathcal{V}_{T^2}^6}\dfrac{e^{4(\alpha-\beta)\phi}}{L^4}\,.
\end{eqaed}
Extremizing in $\phi$ as before now leaves an effective potential,
\begin{eqaed}
    V_{\mathrm{dS}}\,\propto\,\dfrac{\mathcal{V}_{T^2}^2}{N^3}\,,
\end{eqaed}
so $\partial V/\partial\mathcal{V}_{T^2}$ never vanishes, and there is no $\mathrm{dS}$ solution. The runaway is towards small $\mathcal{V}_{T^2}$, where $m^2/V_{\mathrm{dS}}$ grows, so the potential drives the KK mass scale towards $m^2 \gg V$. The Higuchi bound indicated an inconsistency in the naive rigid approximation, and once the moduli are included the would-be extremum turns out to be a runaway direction.
\subsection{Stringy examples}
 
Let us consider Type II supergravity \cite{Polchinski:1998rr, Blumenhagen:2013fgp}  compactified on a 6d group manifold. The ten dimensional string frame action reads, schematically,
\begin{eqaed}\label{eq:10d-action}
    S_{10}=\frac{1}{2\kappa_{10}^2}\int d^{10}x\sqrt{-g_{10}}\left[
    e^{-2\varphi}\left(R_{10}+4(\partial\varphi)^2-\tfrac{1}{2}|H_3|^2\right)
    -\tfrac{1}{2}\sum_q|F_q|^2\right]+S_{CS}+S_{loc},
\end{eqaed}
with $\varphi$ the ten dimensional dilaton, $q$ even in IIA and odd in IIB, and $S_{loc}$ the tension and
charge of the $Dp$-branes and $Op$-planes, the latter entering with negative tension. The ten dimensional string frame metric can be written as
\begin{eqaed}\label{eq:10d-ansatz}
    ds^2_{10}=\tau^{-2} ds^2_{4}+\rho\,d\tilde{s}^2_6,
\end{eqaed}
with $\rho$ the volume modulus of the internal group manifold, and
\begin{eqaed}
    \tau=e^{-\varphi}\rho^{3/2},
\end{eqaed}
the four dimensional dilaton. These are universal moduli, and every supergravity contribution to the potential can be expressed as \cite{Hertzberg:2007wc}
\begin{eqaed}\label{eq:scalingstable}
    V_{H_3}\propto\rho^{-3}\tau^{-2},\quad
    V_{R}=-\tilde{R}\,\rho^{-1}\tau^{-2},\quad
    V_{O6/D6}\propto\tau^{-3},\quad
    V_{F_q}\propto\rho^{3-q}\tau^{-4},
\end{eqaed}
where $\tilde{R}$ is the scalar curvature of $d\tilde{s}^2_6$. In general, we will make no assumption about which combination of fluxes, orientifolds, branes or higher-loop corrections enter the potential, except for demanding that the total potential at the extremum $|V_{*}|$ is not parametrically smaller than $|V_R|$, which we return to below.
 
The Kaluza-Klein masses carry the same scaling as the curvature, since a ten dimensional massless field with eigenvalue $\tilde{m}^2$ of $-\tilde{\Delta}_6$ on $d\tilde{s}^2_6$ scales in four dimensional Einstein frame,
\begin{eqaed}\label{eq:kk-mass-10d}
    m^2_{\rm KK}=\tilde{m}^2\,\rho^{-1}\tau^{-2}.
\end{eqaed}
Comparing \cref{eq:scalingstable} and \cref{eq:kk-mass-10d},
\begin{eqaed}\label{eq:geometric-ratio}
    \dfrac{m^2_{\rm KK}}{V_R}=\dfrac{\tilde{m}^2}{|\tilde{R}|}\qquad\text{for}\qquad \tilde{R}<0,
\end{eqaed}
so that the Higuchi bound $m^2\geq 2 H^2$ \cite{Higuchi:1986py} becomes
\begin{eqaed}\label{eq:higuchi2}
 \dfrac{\tilde{m}^2_{1}}{|\tilde{R}|}\,\geq\,\dfrac{1}{3}\dfrac{V_*}{V_R}.
\end{eqaed}
If the internal curvature is the dominant ingredient in the potential, $V_*\simeq V_R$, this becomes a condition depending only on the underlying algebra and choice of lattice,
\begin{eqaed}\label{eq:higuchi3}
 \tilde{m}^2_{1}\gtrsim |\tilde R|\,.
\end{eqaed}
The conclusion from the previous section carries over straightforwardly to the ten dimensional case, as the ratio in \cref{eq:geometric-ratio} is purely geometric so it is insensitive to how the extremum is constructed. As a concrete example take $(\rm Nil_3)^2$ as internal manifold, where as before
\begin{eqaed}
    \dfrac{m^2_{\rm KK}}{V_R}\simeq \dfrac{1}{\ell_x \ell_y}
\end{eqaed}
so that if the lattice parameters are not dynamical we may take them arbitrarily large and violate the Higuchi bound. As in the previous discussion, this apparent freedom is an artifact of treating $\ell_x,\ell_y$ as parameters rather than fields. A group manifold with vanishing total scalar curvature, for example $\rm Nil_3\times S^3$ with the relative size of the two three-manifolds tuned so that $\tilde R=0$, evades the argument, since the KK masses, $m^2=m^2_{\rm Nil_3}+m^2_{S^3}$, remain finite while $V_R$ vanishes. In this degenerate example with $\tilde R = 0$ the internal curvature has been removed as an ingredient for model building, which is the regime the bound favours.
 
A configuration solving the ten dimensional equations of motion extremizes the four dimensional potential with respect to every modulus of the internal space, since any such deformation is a variation of the ten dimensional fields. The rigid $\rm Nil_3$ example above is in fact not a solution of the $4+3$ dimensional equations of motion, since we only solve the four dimensional ones, and furthermore the Casimir energy does not appear in the classical equations of motion. In particular the left-invariant metric deformations of the group manifold are extremized, so that $\tilde m^2_1$ and $\tilde R$ take definite values on a solution. This is the case for the classical Type IIB de Sitter solutions of \cite{Andriot:2020wpp,Andriot:2020vlg}, which we now look at in some detail\footnote{Note that these solutions are tachyonic, which does not affect our argument.}.

One would ideally evaluate the KK spectrum for these dS solutions to check they satisfy \cref{eq:higuchi2}, however this is quite challenging, as it depends on the precise details of the lattice\footnote{Note that KK gravitons do not receive direct contributions to their mass from internal fluxes or localized internal sources \cite{Martin:2004wp}, unlike scalar modes.}, which is worked out only for two of the solutions \cite{Andriot:2020wpp}.  As a proxy we evaluate the right hand side, which depends only on the curvature as
\begin{eqaed}
    \dfrac{V_*}{V_R}=\dfrac{R_4}{2\,|R_6|}\,,
\end{eqaed}
with $R_4$ and $R_6$ the four and six dimensional curvatures at the solution. We collect these in \Cref{tab:groupmanifolds} for the twenty-two 4d dS solutions of \cite{Andriot:2022yyj}, all of which have $R_6<0$. For these solutions we have on average $V_*/V_R\simeq 10^{-2}$, with all solutions satisfying $V_*/V_R \leq0.2$, so the KK spectrum must satisfy $\tilde m^2_1\gtrsim 10^{-2}|\tilde R|$, two orders of magnitude weaker than \cref{eq:higuchi3}. These classical string constructions therefore sit in the region favoured by the bound. If the moduli on the internal manifold are stabilized such that $\tilde{m}^2_{1}\simeq |\tilde R|$, then these solutions would all satisfy the Higuchi bound.
 \renewcommand{\arraystretch}{1.4}
\begin{table}[H]
\centering
\small
\begin{tabular}{p{3.8cm}  p{2.1cm}  p{2.1cm}  p{2.1cm}  p{2.1cm}}
\hline
algebra & solution & $R_4$ & $|R_6|$ & $V_*/V_R$ \\
\hline
$\mathfrak{iso}(2)\oplus\mathfrak{iso}(2)$   & $s^+_{55}(14)$     & 0.0227    & 0.7577    & 0.0150 \\
\hline
$\mathfrak{iso}(1,1)\oplus\mathfrak{iso}(1,1)$ & $s^+_{55}(15)$     & 0.0194    & 0.8853    & 0.0110 \\
\hline
$\mathfrak{g}^{-1}_{6.76}$                             & $s^+_{55}(16)$     & 0.0498    & 0.8996    & 0.0277 \\
\hline
$\mathfrak{g}^{-1}_{6.76}$                             & $s^+_{55}(17)$     & 0.0568    & 0.8942    & 0.0318 \\
\hline
$\mathfrak{so}(3)\oplus\mathfrak{iso}(1,1)$        & $s^+_{55}(19)$     & 0.0031779 & 0.44861   & 0.0035 \\
\hline
$\mathfrak{so}(3)\oplus\mathrm{Heis}_3$                & $s^+_{55}(20)$     & 0.019450  & 0.72144   & 0.0135 \\
\hline
$\mathfrak{so}(3)\oplus\mathrm{Heis}_3$                & $s^+_{55}(21)$     & 0.023161  & 0.75279   & 0.0154 \\
\hline
$\mathfrak{iso}(1,1)\oplus\mathfrak{iso}(2)$  & $s^+_{55}(22)$     & 0.0028407 & 0.80087   & 0.0018 \\
\hline
$\mathfrak{iso}(1,1)\oplus\mathfrak{iso}(2)$  & $s^+_{55}(23)$     & 0.038665  & 0.77384   & 0.0250 \\
\hline
$\mathfrak{iso}(1,1)\oplus\mathfrak{iso}(2)$  & $s^+_{55}(24)$     & 0.0061178 & 0.79288   & 0.0039 \\
\hline
$\mathfrak{iso}(1,1)\oplus\mathfrak{iso}(2)$  & $s^+_{55}(25)$     & 0.010180  & 0.78633   & 0.0065 \\
\hline
$\mathfrak{iso}(1,1)\oplus\mathfrak{iso}(2)$  & $s^+_{55}(26)$     & 0.026903  & 0.77236   & 0.0174 \\
\hline
$\mathfrak{iso}(1,1)\oplus\mathfrak{iso}(2)$  & $s^+_{55}(27)$     & 0.031484  & 0.77199   & 0.0204 \\
\hline
\hline
$\mathfrak{so}(3)\oplus\mathfrak{so}(3)$               & $s^+_{6666}(1)$    & 0.0019    & 0.4338    & 0.0022 \\
\hline
$\mathfrak{g}^{0,\mu_0,\nu_0}_{6.92}$                  & $s^+_{6666}(3)$    & 0.0066281 & 0.64803   & 0.0051 \\
\hline
$\mathfrak{g}^{0,\mu_0,\nu_0}_{6.92}$                  & $s^+_{6666}(4)$    & 0.033794  & 0.47223   & 0.0358 \\
\hline
\hline
$\mathfrak{so}(3)\oplus \mathfrak{u}(1)^3$              & $m^+_{46}(10)$     & 0.0020327 & 0.041506  & 0.0245 \\
\hline
$\mathfrak{g}^{0,\mu_0,\nu_0}_{6.92}$                  & $m^+_{5577}(3)$    & 0.0045531 & 0.011514  & 0.1977 \\
\hline
$\mathfrak{g}^{0,\mu_0,\nu_0}_{6.92}$                  & $m^+_{5577}(4)$    & 0.011738  & 0.032824  & 0.1788 \\
\hline
$\mathfrak{g}^{0,\mu_0,\nu_0}_{6.92}$                  & $m^+_{5577}(5)$    & 0.0019152 & 0.0071851 & 0.1333 \\
\hline
$\mathfrak{g}^{0,\mu_0,\nu_0}_{6.92}$                  & $m^+_{5577}(6)$    & 0.0023552 & 0.0083805 & 0.1405 \\
\hline
$\mathfrak{g}^{0}_{6.92^*}$                            & $m^{+*}_{5577}(1)$ & 0.0027482 & 0.0092648 & 0.1483 \\
\hline
\end{tabular}
\caption{Classical 4d de Sitter solutions of 10d Type II string theory on compact group manifolds. The solutions are identified in \cite{Andriot:2022yyj} as those admitting compact quotients, and the value of the curvatures at the dS solutions are taken from \cite{Andriot:2020vlg, Andriot:2021rdy, Andriot:2022way}. This follows the classification of \cite{bock2009low}, where $\mathfrak{g}_{d.n}$ denotes the $n$-th $d$-dimensional real unimodular solvable Lie algebra, which includes $\mathfrak{g}^{-1}_{3.4}=\mathfrak{iso}(1,1)$ and $\mathfrak{g}^{0}_{3.5}=\mathfrak{iso}(2)$. Here $\mu_0=-f^{2}_{36},\,\,\nu_0=f^3_{26}$, with $f^{1}_{35}=f^{1}_{24}=1$ and $\mu_0\nu_0\neq0$. Here $\mathfrak{g}^{0}_{6.92^*}$ has $\mu_0=\nu_0$.}
\label{tab:groupmanifolds}
\end{table}
\newpage
\section{Conclusions} \label{sec:conc}

In this work we have proposed a conjecture naturally motivated by the Higuchi bound and the fact that massive spin-2 particles are ubiquitous in theories with compact extra dimensions. Specifically, we have studied whether the KK spectrum associated to a graviton in 4d $\mathrm{dS}$ can somehow violate the bound. In some simple 5d examples we found it cannot, as either one hits another constraint before violating the bound, or the setup simply does not allow for masses below the Hubble scale. In the circle compactification stabilized by Casimir energy, the minimal set of ingredients gives $HR\lesssim 0.13$, and saturating the bound would require shrinking the circle below the species scale. In the warped interval, the warping gaps the tower at $m^2\geq \tfrac{9}{4}H^2$, so KK masses sit above the Hubble scale, independently of $H$ and $L$.

We then argued that internal curvature cannot be the dominant ingredient supporting a $\mathrm{dS}$ extremum, so that group manifolds require additional sources of vacuum energy. The curvature term and the KK masses carry the same power of the volume modulus, so an extremum with $V_*\sim V_R$ sits at $H^2\sim m^2_{\rm KK}$. On $\rm Nil_3$ we found setups that can violate the bound, and showed that they are in fact inconsistent, as the lattice parameters are unstabilized moduli and the volume is a runaway direction. In this sense, requiring that the Higuchi bound is not violated by smooth deformations is a diagnostic tool for whether an effective theory is consistent or not, and one that only requires the KK spectrum and the curvature. We then looked at Type IIB solutions to the 10d equations of motion found in the literature, and argued that they are precisely in the region favoured by the Higuchi bound.

Some questions remain open. In the strongly warped regime the species scale no longer matches the five dimensional Planck mass, suggesting that the standard species counting and the emergent string conjecture may require modification away from the strict Minkowski limit. It would also be interesting to investigate whether further stabilization of the interval selects a particular relation between $H$ and $L$, and therefore a particular scaling of the tower with the vacuum energy. For instance, the dark dimension scaling $L\simeq \Lambda_{cc}^{-1/4}$ can be reproduced in the warped interval, but since $L$ is a free parameter this has to be imposed as an additional assumption. In a different direction, a tuned cancellation leaving the vacuum energy parametrically below the curvature term evades our argument, and whether such a cancellation can arise in a controlled realization remains to be understood. Computing the KK spectrum for the solutions of \Cref{tab:groupmanifolds} would test the bound directly, though this requires the full lattice data and is only tractable for the simplest internal manifolds.
 Finally, while a bound on KK graviton masses follows directly from the Higuchi bound, it is not a priori restricting light towers with spin $s<2$. The cosmological implications of a tower of light scalars with masses $m\ll H$ were studied in \cite{Lust:2025auk}, with the assumption there was no accompanying light tower of massive gravitons. It has been shown that warped compactification can allow for a spin dependent mass gap, but whether a setup with a hierarchy of the form $m_{\text{KK},s=0}^2<2H^2<m_{\text{KK},s=2}^2$ is consistent in string theory, or more generally in higher-dimensional gravity, is a question for future work.
\section*{Acknowledgments}
We would like to thank Luis Anchordoqui, David Andriot, Alvaro Herráez, Daniel Junghans and Marco Scalisi for stimulating conversations and comments on this draft. The work of DL is supported by the German-Israel-Project (DIP) on Holography and the Swampland.
\appendix
\section{KK masses on a warped background}
\label{sec:kkwarp}
We consider the warped background
\begin{equation}
ds^2 = dy^2 + A(y)^2\,\bar g_{ij}(x)\,dx^i dx^j,
\end{equation}
where the $d$-dimensional metric $\bar g_{ij}$ is Einstein
\begin{equation}
\bar R_{ij} = \dfrac{\bar R}{d}\,\bar g_{ij},
\end{equation}
with $\beta$ a constant. For de Sitter slices one has $\bar R=12 H^2$.

We are interested in the spin-2 perturbations of the lower dimensional metric on the slices,
\begin{equation}
G_{ij} = A(y)^2\bigl(\bar g_{ij} + h_{ij}(x,y)\bigr),\qquad
G_{iy}=0,\qquad G_{yy}=1,
\end{equation}
and impose the transverse-traceless conditions on the slice,
\begin{equation}
\bar\nabla^i h_{ij}=0,
\qquad
\bar g^{ij}h_{ij}=0.
\end{equation}
This can be written as
\begin{equation}
\nabla^M(A(y)^2 h_{MN})=0, \qquad G^{MN}(A(y)^2 h_{MN})=0.
\end{equation}
We then use the standard result \cite{Wald:1984rg}
\begin{align}
    R[g+\delta G]_{\mu\nu}=R[g]_{\mu\nu}+ \frac{1}{2}\left(-\nabla^2 \delta G_{\mu\nu} - \nabla_\mu\nabla_\nu \delta G + \nabla_\mu\nabla_\rho \delta G^\rho{}_\nu + \nabla_\nu\nabla_\rho \delta G^\rho{}_\mu\right)\\ 
    - R_{\mu\rho\nu\sigma}\delta G^{\rho\sigma} + R_{\rho(\mu}\delta G^\rho{}_{\nu)},
\end{align}
which for TT modes simplifies as
\begin{equation}
    R[G+\delta G]_{\mu\nu}=R[g]_{\mu\nu}- \frac{1}{2}\left(\nabla^2 \delta G_{\mu\nu}\right) - R_{\mu\rho\nu\sigma}\delta G^{\rho\sigma} + R_{\rho(\mu}\delta G^\rho{}_{\nu)}.
\end{equation}
Then, for $\delta G_{ij}=A(y)^2 h_{ij}$, we can compute the linearized Einstein equations, which simplify as
\begin{equation}
A^2 h_{ij}'' + dA'A\, h_{ij}' + \bar\nabla^2 h_{ij}
+2\bar R_{ikjl}h^{kl} = 0,
\end{equation}
where we have used both the TT condition and the fact that the background is Einstein, and everything is now with respect to the $\bar g_{ij}$ metric.

It is convenient to rewrite this in terms of the Lichnerowicz operator
\begin{equation}
(\Delta_L^{(\bar g)} h)_{ij}
=
-\bar\nabla^2 h_{ij}
-2\bar R_{ikjl}h^{kl}
+2\bar R_{k(i}h^k{}_{j)}.
\end{equation}
Since the slice is Einstein, we can rearrange this as
\begin{equation}
\bar\nabla^2 h_{ij}+2\bar R_{ikjl}h^{kl}
=
-\bigl(\Delta_L^{(\bar g)}h\bigr)_{ij}
+\dfrac{2}{d}\bar R\,h_{ij}.
\end{equation}

Finally, we see that the helicity-2 mode for the lower dimensional graviton has equation of motion
\begin{equation}
A^2 h_{ij}'' + dA'A\, h_{ij}'
-\bigl(\Delta_L^{(\bar g)}-\dfrac{2}{d}\bar R\bigr)h_{ij}=0.
\end{equation}
This is enough to determine the KK masses. As a sanity check we look at the zero mode, defined by vanishing Fierz-Pauli mass,
\begin{equation}
m^2=0.
\end{equation}
In this case the equation on the $y$ dependent part becomes
\begin{equation}
\psi_0'' + d\frac{A'}{A}\psi_0' = 0.
\end{equation}
This has solution
\begin{equation}
A^d\psi_0' = C
\end{equation}
for some constant $C$. If we impose Neumann boundary conditions at the endpoints of the interval,
\begin{equation}
\psi_0'(y)\big\vert_{y=0,L}=0,
\end{equation}
then necessarily
\begin{equation}
C=0,
\end{equation}
and
\begin{equation}
\psi_0'(y)=0.
\end{equation}
Thus the zero-mode profile is constant,
\begin{equation}
\psi_0(y)=\text{constant}.
\end{equation}

The massless lower dimensional graviton then satisfies
\begin{equation}
\bigl(\Delta_L^{(\bar g)}-\dfrac{2}{d}\bar R\bigr)h^{(0)}_{ij}=0.
\end{equation}
This is precisely the massless spin-$2$ equation on the Einstein slice, so indeed the zero mode is a massless graviton with constant profile along the interval.
\subsection{Generalization to \texorpdfstring{$p$}{p} compact directions}
Let us consider a slight generalization, for a $d+p$ dimensional metric of the form 
\begin{equation}
ds^2 =A(y)^2\,\bar g_{ij}(x)\,dx^i dx^j+ g_{mn}(y)\,dy^m dy^n.
\end{equation}
Assuming again Einstein slices the generalization is straightforward \cite{Bachas:2011xa},
\begin{equation}
A^2\left[\nabla^{(g)}_p \nabla_{(g)}^p h_{ij} + d\,\frac{\nabla^{(g)}_p A}{A}\,\nabla_{(g)}^p h_{ij}\right] - \bigl(\Delta_L^{(\bar g)} - \dfrac{2}{d}\bar R\bigr)h_{ij} = 0.
\end{equation}

\section{Casimir energy on \texorpdfstring{$\rm Nil_3$}{Nil3}}
\label{sec:nil3casimir}

In this appendix we evaluate the Casimir contribution to the scalar potential from the twisted sector of the $\rm Nil_3$ KK spectrum, quoted in the main text. Recall the mass spectrum and degeneracy of the twisted modes,
\begin{eqaed}
m^2_{n,k}=\dfrac{2 N \pi}{\ell_x \ell_y}\left(k (2n+1)+\frac{2 \pi  k^2 N}{\ell_x \ell_y}\right),
\qquad
d_{n,k}=k N,
\end{eqaed}
with $n\geq 0$ and $k\geq 1$, and the potential
\begin{equation}
  V \,=\, -\frac{1}{2\,(4\pi)^{d/2}} \int_0^\infty \mathrm{d}t\,
  t^{-1-\frac{d}{2}}\, K(t),
  \qquad
  K(t) \,=\, \sum_{n,k} d_{n,k}\, e^{-t m^2_{n,k}} .
\end{equation}
We first perform the sum in $n$. Writing $\lambda_z=k\lambda_0$ with $\lambda_0=\frac{2\pi N}{\ell_x \ell_y}$, so that $m^2_{n,k}=\lambda_z^2+(2n+1)|\lambda_z|$, the sum at fixed $k$ is geometric,
\begin{eqaed}
    \sum_{n\geq 0}e^{-t(2n+1)|\lambda_z|}=\frac{e^{-t|\lambda_z|}}{1-e^{-2t|\lambda_z|}}=\frac{1}{2\sinh\left(t|\lambda_z|\right)},
\end{eqaed}
and
\begin{eqaed}
    K(t)=\sum_{k\neq 0}|k|N\,e^{-t\lambda_z^2}\sum_{n\geq 0}e^{-t(2n+1)|\lambda_z|}=N\sum_{k\geq 1}\frac{k\,e^{-t\lambda_0^2k^2}}{\sinh(t\lambda_0 k)}.
\end{eqaed}
Rescaling $t\to\tau/\lambda_0$ gives
\begin{eqaed}
  V = -\left(\frac{2\pi N}{\ell_x\,\ell_y}\right)^{\frac{d}{2}}\frac{N}{2\,(4\pi)^{d/2}} \int_0^\infty \mathrm{d}\tau\,
  \tau^{-1-\frac{d}{2}} \sum_{k\geq 1}   k \,\dfrac{e^{-\lambda_0 \tau k^2}}{\sinh{k \tau}},
\end{eqaed}
so the integral has to be computed via zeta function regularization. For $\lambda_0\ll 1$ the gaussian can be set to 1 at leading order\footnote{We check numerically that this is a good approximation for $\lambda_0\lesssim0.01$, with relative error $<1\%$.}, and expanding $1/\sinh u=2\sum_{m\geq 0}e^{-(2m+1)u}$ the $\tau$ integral can be performed analytically, followed by the sums in $m$ and $k$. We find at leading order in $\lambda_0$
\begin{eqaed}
    V=-\frac{N\lambda_0^{d/2}}{(4\pi)^{d/2}}\Gamma(-\frac{d}{2})\zeta(-\frac{d}{2})\zeta(-\frac{d}{2}-1)\left(1-2^{\frac{d}{2}}\right).
\end{eqaed}
We can now evaluate this using 
\begin{eqaed}
    \Gamma(-2n)\zeta(-2n)=
(-1)^n \frac{\zeta(2n+1)}{2^{2n+1}\pi^{2n}},
\end{eqaed}
and we obtain 
\begin{eqaed}
     V_{d=4} \,=\, -\frac{\zeta(3)}{5120\,\pi^4}N\lambda_0^2\,=\,-\frac{\zeta(3)}{1280\,\pi^2}\dfrac{N^3}{(\ell_x \ell_y)^2}\,\simeq\,-9.5\times 10^{-5}\dfrac{N^3}{(\ell_x \ell_y)^2},
\end{eqaed}
which is the result quoted in the main text.
\printbibliography
\end{document}